\documentclass[aps,prd,twocolumn,superscriptaddress,showpacs,preprintnumbers,amsmath,amssymb]{revtex4}

\usepackage{graphicx}
\graphicspath{{ps/}}
\usepackage{dcolumn}

\begin{document}


\title{\quad\\[1.0cm] Study of the decays
{\boldmath $B\to D_{s1}(2536)^+\bar D^{(*)}$}}
\affiliation{Budker Institute of Nuclear Physics, Novosibirsk}
\affiliation{Faculty of Mathematics and Physics, Charles University, Prague}
\affiliation{University of Cincinnati, Cincinnati, Ohio 45221}
\affiliation{Department of Physics, Fu Jen Catholic University, Taipei}
\affiliation{Justus-Liebig-Universit\"at Gie\ss{}en, Gie\ss{}en}
\affiliation{Gifu University, Gifu}
\affiliation{The Graduate University for Advanced Studies, Hayama}
\affiliation{Hanyang University, Seoul}
\affiliation{University of Hawaii, Honolulu, Hawaii 96822}
\affiliation{High Energy Accelerator Research Organization (KEK), Tsukuba}
\affiliation{University of Illinois at Urbana-Champaign, Urbana, Illinois 61801}
\affiliation{Institute of High Energy Physics, Chinese Academy of Sciences, Beijing}
\affiliation{Institute of High Energy Physics, Vienna}
\affiliation{Institute of High Energy Physics, Protvino}
\affiliation{Institute for Theoretical and Experimental Physics, Moscow}
\affiliation{J. Stefan Institute, Ljubljana}
\affiliation{Kanagawa University, Yokohama}
\affiliation{Institut f\"ur Experimentelle Kernphysik, Karlsruher Institut f\"ur Technologie, Karlsruhe}
\affiliation{Korea Institute of Science and Technology Information, Daejeon}
\affiliation{Korea University, Seoul}
\affiliation{Kyungpook National University, Taegu}
\affiliation{\'Ecole Polytechnique F\'ed\'erale de Lausanne (EPFL), Lausanne}
\affiliation{Faculty of Mathematics and Physics, University of Ljubljana, Ljubljana}
\affiliation{University of Maribor, Maribor}
\affiliation{Max-Planck-Institut f\"ur Physik, M\"unchen}
\affiliation{University of Melbourne, School of Physics, Victoria 3010}
\affiliation{Nagoya University, Nagoya}
\affiliation{Nara Women's University, Nara}
\affiliation{National Central University, Chung-li}
\affiliation{National United University, Miao Li}
\affiliation{Department of Physics, National Taiwan University, Taipei}
\affiliation{H. Niewodniczanski Institute of Nuclear Physics, Krakow}
\affiliation{Nippon Dental University, Niigata}
\affiliation{Niigata University, Niigata}
\affiliation{University of Nova Gorica, Nova Gorica}
\affiliation{Novosibirsk State University, Novosibirsk}
\affiliation{Osaka City University, Osaka}
\affiliation{Panjab University, Chandigarh}
\affiliation{Saga University, Saga}
\affiliation{University of Science and Technology of China, Hefei}
\affiliation{Seoul National University, Seoul}
\affiliation{Sungkyunkwan University, Suwon}
\affiliation{School of Physics, University of Sydney, NSW 2006}
\affiliation{Tata Institute of Fundamental Research, Mumbai}
\affiliation{Excellence Cluster Universe, Technische Universit\"at M\"unchen, Garching}
\affiliation{Toho University, Funabashi}
\affiliation{Tohoku Gakuin University, Tagajo}
\affiliation{Tohoku University, Sendai}
\affiliation{Tokyo Institute of Technology, Tokyo}
\affiliation{Tokyo Metropolitan University, Tokyo}
\affiliation{Tokyo University of Agriculture and Technology, Tokyo}
\affiliation{CNP, Virginia Polytechnic Institute and State University, Blacksburg, Virginia 24061}
\affiliation{Wayne State University, Detroit, Michigan 48202}
\affiliation{Yonsei University, Seoul}
  \author{T.~Aushev}\affiliation{Institute for Theoretical and Experimental Physics, Moscow} 
  \author{I.~Adachi}\affiliation{High Energy Accelerator Research Organization (KEK), Tsukuba} 
  \author{K.~Arinstein}\affiliation{Budker Institute of Nuclear Physics, Novosibirsk}\affiliation{Novosibirsk State University, Novosibirsk} 
  \author{V.~Aulchenko}\affiliation{Budker Institute of Nuclear Physics, Novosibirsk}\affiliation{Novosibirsk State University, Novosibirsk} 
  \author{A.~M.~Bakich}\affiliation{School of Physics, University of Sydney, NSW 2006} 
  \author{V.~Balagura}\affiliation{Institute for Theoretical and Experimental Physics, Moscow} 
  \author{V.~Bhardwaj}\affiliation{Panjab University, Chandigarh} 
  \author{M.~Bischofberger}\affiliation{Nara Women's University, Nara} 
  \author{A.~Bondar}\affiliation{Budker Institute of Nuclear Physics, Novosibirsk}\affiliation{Novosibirsk State University, Novosibirsk} 
  \author{A.~Bozek}\affiliation{H. Niewodniczanski Institute of Nuclear Physics, Krakow} 
% ...
\collaboration{The Belle Collaboration}

\begin{abstract}
We report a study of the decays $B\to D_{s1}(2536)^+\bar D^{(*)}$,
where $\bar D^{(*)}$ is $\bar D^0$, $D^-$ or $D^{*-}$, using a sample
of $657\times10^6$ $B\bar{B}$ pairs collected at the $\Upsilon(4S)$
resonance with the Belle detector at the KEKB asymmetric-energy
$e^+e^-$ collider.  The branching fractions of the decays $B^+\to
D_{s1}(2536)^+\bar D^0$, $B^0\to D_{s1}(2536)^+D^-$ and $B^0\to
D_{s1}(2536)^+D^{*-}$ multiplied by that of $D_{s1}(2536)^+\to
(D^{*0}K^++D^{*+}K^0)$ are found to be
$(3.97\pm0.85\pm0.56)\times10^{-4}$,
$(2.75\pm0.62\pm0.36)\times10^{-4}$ and
$(5.01\pm1.21\pm0.70)\times10^{-4}$, respectively.
The ratio ${\cal B}(D_{s1}\to D^{*0}K^+)/{\cal B}(D_{s1}\to D^{*+}K^0)$ 
is measured to be $0.88\pm0.24\pm0.08$.
\end{abstract}

\maketitle

The $D_{s1}(2536)^+$ is a narrow P-wave resonance for which the
$J^P=1^+$ assignment is strongly favored~\cite{pdg10}.  Although the
$D_{s1}(2536)^+$ was first observed in 1989~\cite{dsj1_observ}, its
properties are still not well measured.  In this analysis we study the
production of $D_{s1}(2536)^+$ in doubly charmed $B$ meson decays,
$B\to D_{s1}(2536)^+\bar D^{(*)}$, where $\bar D^{(*)}$ is either a
$\bar D^0$, $D^-$ or $D^{*-}$~\cite{cc}.  The branching fraction measurements of
the decays $B\to D_{s1}(2536)^+\bar D^{(*)}$ together with those of
$B\to\bar D^{(*)}D^{(*)}_{s(J)}$ decays provide important information to
check the molecular hypothesis for the $D^*_{s0}(2317)$ and 
$D_{s1}(2460)$ particles~\cite{orsay,datta}.  First observations of 
the $B\to D_{s1}(2536)^+\bar D^{(*)}$ decay modes have been reported 
by BaBar~\cite{dstdstks_babar,ddk_babar}.  An upper limit on the decay
$B^0\to D_{s1}(2536)^+D^{*-}$ was also obtained by
Belle~\cite{dstdstks_belle}, which is consistent with the BaBar
measurement.

This analysis is based on $605\,$fb$^{-1}$ of data collected at the
$\Upsilon(4S)$ resonance with the Belle detector~\cite{belle} at the
KEKB asymmetric-energy $e^+e^-$ collider~\cite{KEKB}, which
corresponds to $657\times10^6$ $B\bar B$ pairs.  The Belle detector is
a general-purpose spectrometer with a $1.5{\rm~T}$ magnetic field
provided by a superconducting solenoid.  A silicon vertex detector and
a 50-layer central drift chamber are used to measure the momenta of
charged particles.  Photons are detected in an electromagnetic
calorimeter consisting of CsI(Tl) crystals.  Particle
identification likelihoods ${\cal L}_K$ and ${\cal L}_{\pi}$ are
derived from information provided by an array of time-of-flight
counters, an array of silica aerogel Cherenkov threshold counters and
$dE/dx$ measurements in the central drift chamber.  Two
inner detector configurations were used.  A 2.0~cm radius beampipe and
a 3-layer silicon vertex detector were used for the first sample of
$152\times10^6$ $B\bar B$ pairs, while a 1.5~cm radius beampipe, a
4-layer silicon detector and a small-cell inner drift chamber were
used to record the remaining $505\times10^6$ $B\bar B$ pairs.

All charged tracks are required to have a distance of closest approach
to the interaction point (IP) in the plane perpendicular to the beam
axis smaller than $2\,$cm and smaller than $5\,$cm along the beam 
axis.  Charged kaon and pion candidates are required to be
positively identified.  The $K_S^0$ candidates are reconstructed in
the $\pi^+\pi^-$ mode with the requirement
$|M_{\pi\pi}-m_{K_S^0}|<15{\rm~MeV}/c^2$ ($3\sigma$), where
$m_{K_S^0}$ is the $K_S^0$ mass~\cite{pdg10}.  Requirements on the
$K_S^0$ vertex displacement from the IP and on the difference between
the vertex and $K_S^0$ flight directions are applied.  No pion
identification is required for the pions from $K_S^0$ candidates.  
A mass- and vertex-constrained fit is applied to improve the
four-momentum measurements of $K_S^0$ candidates.
Photons are reconstructed in the electromagnetic calorimeter from showers that are not
associated with charged tracks with energies larger than 50~MeV.
Combinations of two photons are considered to be $\pi^0$ candidates if
their invariant mass lies within $\pm15{\rm~MeV}/c^2$ ($3\sigma$) of
the $\pi^0$ mass~\cite{pdg10}.  To improve their momentum resolution,
all $\pi^0$ candidates are fitted with a $\pi^0$ mass constraint.
Continuum $e^+e^-\to q\bar q$ backgrounds ($q=u,d,s,c)$ are suppressed
by requiring the ratio of the second and zeroth Fox-Wolfram
moments~\cite{fox} to be smaller than $0.3$.

Candidate $D^0$'s are reconstructed using five decay modes:
$K^-\pi^+$, $K^-\pi^+\pi^+\pi^-$, $K^-\pi^+\pi^0$, $K_S^0\pi^+\pi^-$
and $K^+K^-$.  The $D^-$ is reconstructed via its decay into
$K^+\pi^-\pi^-$.  The selected combinations are constrained to a
common vertex and the $\chi^2/n.d.f.$ of the vertex fit is required to
be smaller than $25$.  A $\pm15{\rm~MeV}/c^2$ ($2\sigma$) mass window
around the $D^0$ mass~\cite{pdg10} is used to select $D^0$ candidates
for the $D^0\to K^-\pi^+\pi^0$ mode and $\pm12{\rm~MeV}/c^2$
($\approx3\sigma$) for the other $D$ decay modes.  To reduce the large
combinatorial background in the $D^0\to K^-\pi^+\pi^0$ decay mode, the
energy of each photon from $\pi^0$ is required to be greater than
100~MeV.  For the $\bar D^0$'s coming directly from $B$ decay, 
only the cleanest modes, $K^+\pi^-$, $K_S^0\pi^+\pi^-$ and $K^+K^-$,
are used.  A mass- and vertex-constrained fit is applied for all $D$
candidates.
The $D^{*+}$ is reconstructed via its decay to $D^0\pi_{\rm slow}^+$.
To improve the $\pi_{\rm slow}^+$ momentum resolution, and thus the
$D^{*+}$ mass resolution, the $\pi_{\rm slow}^+$ is constrained to the
$D^{*+}$ decay vertex, which is obtained by fitting the $D^0$ to the
IP profile.  The $D^{*+}$ candidates with invariant mass within
$\pm2{\rm~MeV}/c^2$ ($4\sigma$) of the $D^{*+}$~\cite{pdg10} mass are
selected.
The $D^{*0}$ is reconstructed using the $D^0\pi^0$ and $D^0\gamma$
decay modes with invariant mass windows of $\pm3{\rm~MeV}/c^2$ and
$\pm12{\rm~MeV}/c^2$ ($2\sigma$), respectively.  In both cases a
mass-constrained fit is performed for $D^{*0}$ candidates.

\begin{figure*}[htbp]
  \begin{center}
  \includegraphics[width=.32\textwidth]{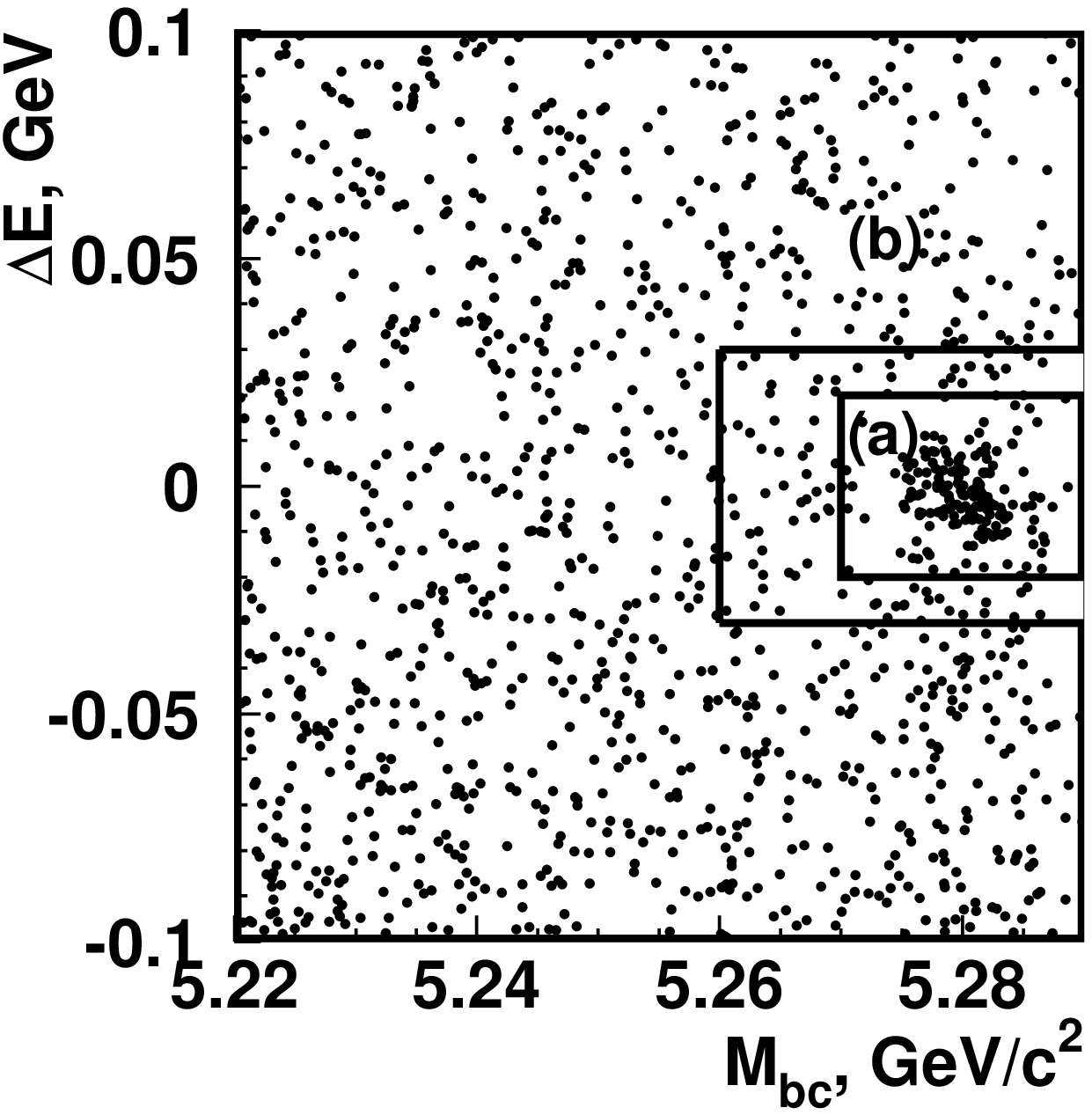}
  \includegraphics[width=.32\textwidth]{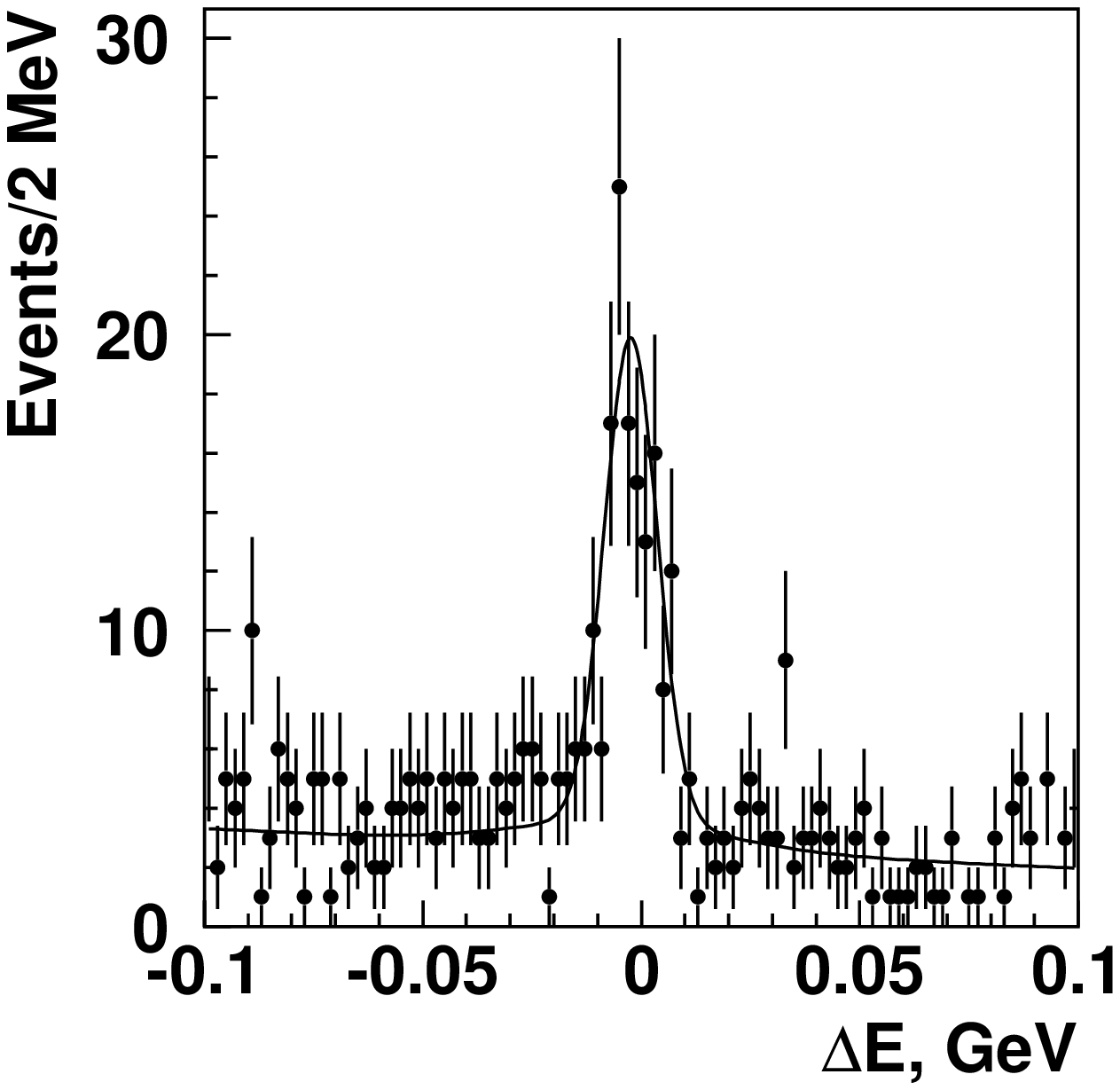}
  \includegraphics[width=.32\textwidth]{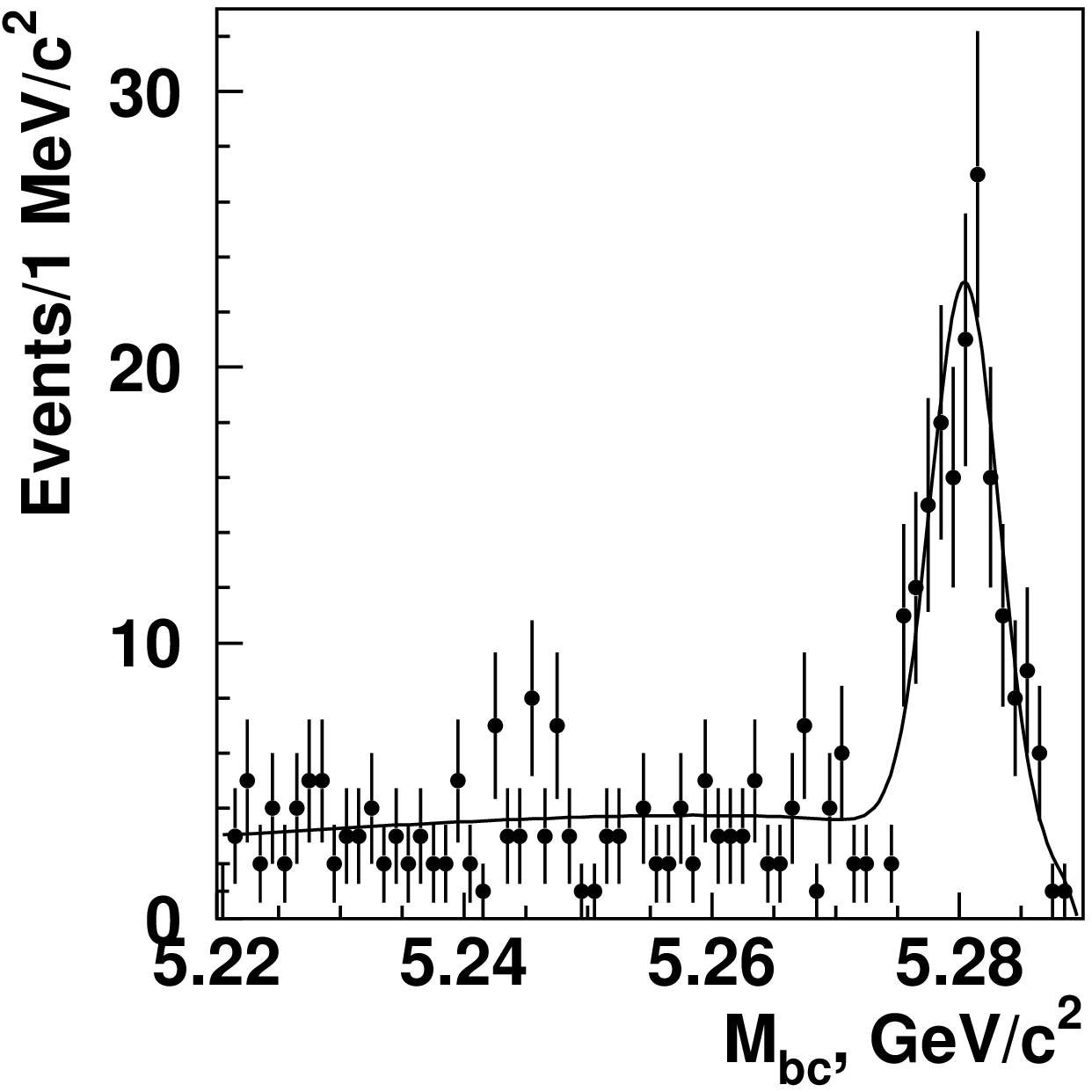}
  \caption{The scatter plot of $\Delta E$ vs $M_{\rm bc}$ (left), 
  	where (a) is the signal region and (b) is the sideband, and 
	projections on $\Delta E$ for events with $M_{\rm bc}>5.27
	{\rm~GeV}/c^2$ (center) and $M_{\rm bc}$ for events with $|\Delta
    E|<0.02{\rm~GeV}$ (right).  The $M_{\rm bc}$ distribution is
    parametrized with a single Gaussian for the signal and an ARGUS
    function~\cite{argusfunc} for the background.  The $\Delta E$ is
    parametrized with a double Gaussian and a linear polynomial for
    the signal and background, respectively.  The results of the fits
    are shown by the superimposed curves.}
  \label{de_mbc_data}
  \end{center}
\end{figure*}

The $D_{s1}(2536)^+$ meson is reconstructed in its dominant decay
modes: $D^{*0}K^+$ and $D^{*+}K_S^0$.  The invariant mass of the
$D_{s1}(2536)^+$ candidates is required to be less than
$2.58{\rm~GeV}/c^2$.
Combinations of $D_{s1}(2536)^+$ and a second charm $\bar D^{(*)}$
meson ($\bar D^{(*)} = \bar D^0$, $D^-$, $D^{*-}$) with opposite charm
flavor are considered as $B$ meson candidates.
$B$ candidates are identified using the energy difference $\Delta
E=E_B-E_{\rm beam}$ and beam-energy constrained mass $M_{\rm
  bc}=\sqrt{E_{\rm beam}^2-p_B^{*2}}$, where $E_B$ and $p_B^*$ are the
$B$ candidate energy and momentum in the center-of-mass system,
respectively.  We require $|\Delta E|<0.1\,$GeV and $M_{\rm 
  bc}>5.22{\rm~GeV}/c^2$ to preselect $B$ candidates.  After these
selections the average candidate multiplicity per event is $1.6$.  A
single candidate per event is chosen based on the smallest $\chi^2$ of
the $D^{(*)}$ candidates mass deviations from their nominal values.
The $\Delta E$ versus $M_{\rm bc}$ scatter plot for the candidates
with $D_{s1}(2536)^+$ mass within $\pm5{\rm~MeV}/c^2$ ($3\sigma$) of
the nominal value~\cite{pdg10} is presented in Fig.~\ref{de_mbc_data}
left, for the sum of all studied decay modes.  A signal box is defined
as $|\Delta E|<0.02$~GeV and $M_{\rm bc}>5.27{\rm~GeV}/c^2$ (region
``a'').  The $\Delta E-M_{\rm bc}$ two-dimensional sideband is defined
as $M_{\rm bc}<5.26{\rm~GeV}/c^2$ or $|\Delta E|>0.03{\rm~GeV}$
(region ``b'').  The $M_{\rm bc}$ ($\Delta E$) projection is shown for
the events with $|\Delta E|<0.02$~GeV ($M_{\rm bc}>5.27{\rm~GeV}/c^2$)
in Fig.~\ref{de_mbc_data} center (right).

\begin{figure*}[htb]
  \includegraphics[width=.89\textwidth]{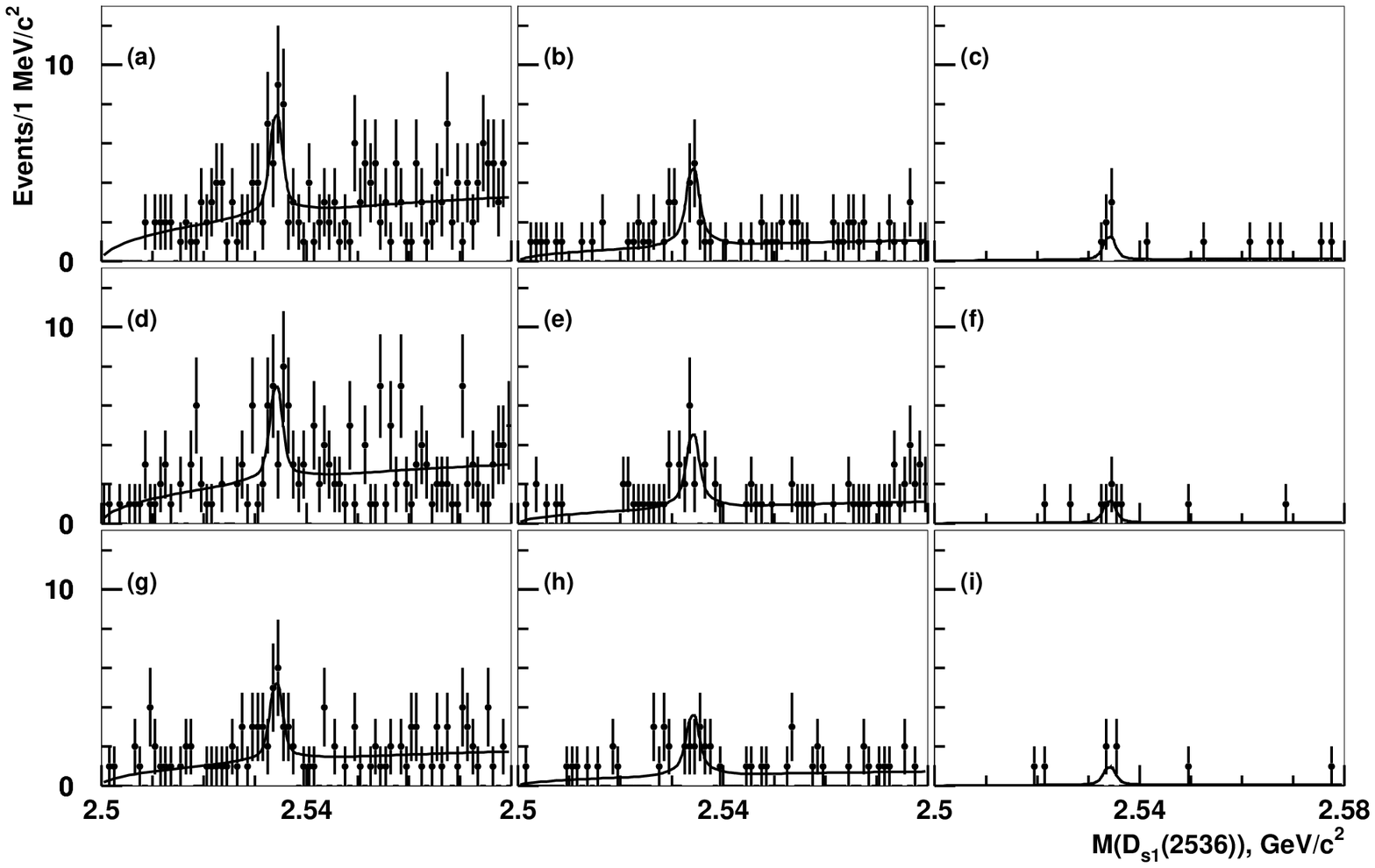}
  \caption{
    $D_{s1}(2536)^+$ mass distributions for:
    a), b), c) $B^+\to D_{s1}(2536)^+\bar D^0$;
    d), e), f) $B^0\to D_{s1}(2536)^+ D^-$ and
    g), h), i) $B^0\to D_{s1}(2536)^+ D^{*-}$ final states,
    followed by $D_{s1}(2536)^+$ decays to
    a), d), g) $D^{*0}(D^0\gamma)K^+$;
    b), e), h) $D^{*0}(D^0\pi^0)K^+$ and
    c), f), i) $D^{*+}(D^0\pi^+)K_S^0$.
    The points with error bars are the data, while the curves 
    show the fit result.
  }
  \label{mds2536_data}
\end{figure*}

A two-dimensional binned likelihood fit is performed to the $\Delta
E-M_{\rm bc}$ distribution and yields $151\pm15$ signal events.  The
signal yield obtained from the $\Delta E-M_{\rm bc}$ fit contains a
contribution from non-resonant $B\to D^*K\bar D^{(*)}$ decays that 
have the same final state particles as the signal.  Therefore to 
extract the yields of each decay mode, we perform an additional binned 
likelihood fit to the $D_{s1}(2536)^+$ mass distributions for the 
events in the $\Delta E-M_{\rm bc}$ signal box 
(Fig.~\ref{mds2536_data}).  In total, there are nine reconstructed 
decay modes corresponding to three $B$ meson decay modes, $B^+\to 
  D_{s1}(2536)^+\bar D^0$, $B^0\to D_{s1}(2536)^+D^-$ and $B^0\to 
  D_{s1}(2536)^+D^{*-}$, and three $D_{s1}(2536)$ meson decay modes, 
$D_{s1}(2536)^+\to D^{*0}(D^0\gamma)K^+$, $D^{*0}(D^0\pi^0)K^+$ and
$D^{*+}(D^0\pi^+)K_S^0$.  The latter are related by the known 
branching ratios $R^\prime={\cal B}(D^{*0}\to D^0\pi^0)/{\cal
  B}(D^{*0}\to D^0\gamma)=1.74\pm0.13$ and $R^{\prime\prime}={\cal
  B}(D_{s1}(2536)^+\to D^{*0}K^+)/{\cal B}(D_{s1}(2536)^+\to
D^{*+}K^0)=1.36\pm0.20$~\cite{pdg10}.  The shapes of the signal
distributions are taken to be a non-relativistic Breit-Wigner functions
convolved with the $D_{s1}(2536)^+$ mass resolution function.  The
Breit-Wigner mass and width of $D_{s1}(2536)^+$ are floating
parameters common to all modes.  The $D_{s1}(2536)^+$ mass resolution
is parametrized by a double Gaussian; the resolution parameters and
reconstruction efficiencies are obtained from Monte Carlo (MC)
simulation and are summarized in Table~\ref{tableres}.
\begin{table}[htb]
\caption{The products of the total efficiencies and intermediate
  branching fractions, $\epsilon{\cal B}$, and the $D_{s1}(2536)^+$
  mass resolution parameters, where $\sigma_{\rm main}$ ($\sigma_{\rm
    tail}$) is the width of the narrow (wide) Gaussian component, and 
    $f_{\rm tail}$ is the fraction of the wide component.}
\begin{center}
\begin{tabular}{lccc}
\hline
$D_{s1}$ mode: & 
		$D^{*0}(D^0\gamma)K^+$&$D^{*0}(D^0\pi^0)K^+$&$D^{*+}K_S$\\
\hline
$\epsilon{\cal B}(D_{s1}\bar D^0),\,10^{-4}$&$1.45$&$1.18$&$0.74$\\
$\epsilon{\cal B}(D_{s1}D^-),     \,10^{-4}$&$1.97$&$1.62$&$1.00$\\
$\epsilon{\cal B}(D_{s1}D^{*-}),  \,10^{-4}$&$0.93$&$0.75$&$0.32$\\
\hline
$\sigma_{\rm main}$, MeV/$c^2$ & 1.02 & 1.01 & 0.95 \\
$\sigma_{\rm tail}$, MeV/$c^2$ & 5.44 & 3.61 & 2.43 \\
$f_{\rm tail}$                 & 0.23 & 0.34 & 0.15 \\
\hline
\end{tabular}
\end{center}
\label{tableres}
\end{table}
The shape of the background is parametrized with an ARGUS
function~\cite{argusfunc} and fixed from a fit to events in the $\Delta E-M_{\rm bc}$
sideband.  This background function includes both the non-resonant
component and the combinatorial background since their shapes are
expected to be similar.  All nine distributions are fitted
simultaneously.  The ratios of signal yields in
different $D_{s1}(2536)^+$ decay modes are fixed using their relative
branching fractions, $R^\prime$ and $R^{\prime\prime}$, and MC
reconstruction efficiencies.  The number of signal events for each of
the $B$ decay modes is floated in the fit as a sum of the signal
events in all $D_{s1}(2536)^+$ decay modes; the background
normalization is free for each distribution.  The results of the fit
are presented in Table~\ref{results}.  The significance is calculated
from ${\cal S}=\sqrt{-2\ln({\cal L}_{0}/{\cal L}_{\mathrm{max}})}$,
where ${\cal L}_{0}$ and ${\cal L}_{\mathrm{max}}$ are the maximized
likelihoods with the signal yield fixed at zero and left free,
respectively.  The $D_{s1}(2536)^+$ mass is found to be
$2534.1\pm0.6{\rm~MeV}/c^2$ and the width is
$\Gamma=0.75\pm0.23{\rm~MeV}/c^2$, consistent with the
current world-average values in~\cite{pdg10}.  To provide a measurement of the
$D_{s1}(2536)^+$ branching-fractions ratio 
another fit is performed with a floating $R^{\prime\prime}$ parameter.
From the fit we obtain $R^{\prime\prime}=0.88\pm0.24$, consistent with 
other measurements~\cite{pdg10}.

\begin{table}[htb]
\caption{Summary of the fit results: event yields, branching
  fractions, and statistical significances.}
\begin{center}
\begin{tabular}{lccc}
\hline
$B$ decay mode & $N$ & ${\cal B},\,10^{-4}$ & $\cal S$ \\
\hline
$D_{s1}(2536)^+(D^*K)\bar D^0$
	& $42.5\pm8.7$ & $\,\,3.97\pm0.85\,\,$ & $7.0\sigma$ \\
$D_{s1}(2536)^+(D^*K)D^-$
	& $40.2\pm8.6$ & $2.75\pm0.62$ & $6.9\sigma$ \\
$D_{s1}(2536)^+(D^*K)D^{*-}$
	& $33.3\pm7.6$ & $5.01\pm1.21$ & $6.3\sigma$ \\
\hline
\end{tabular}
\end{center}
\label{results}
\end{table}

The main systematics for the branching fraction measurements is from
the tracking efficiency.  A $1\%$ systematic error is assigned for
each track and an additional $3\%$ error for each low momentum track,
which are summed linearly.  For the kaon identification a $1\%$
systematic uncertainty is assigned for each kaon track.  The 
contributions of the systematic uncertainties of the $K_S^0$ and 
$\gamma/\pi^0$ reconstruction efficiencies to the overall systematics 
are estimated to be $1\%$ and $3\%$, respectively.  The width of the 
narrow Gaussian component of the signal mass resolution is increased 
by 20\% to obtain the systematics due to the poorer resolution in data 
compared to MC simulation.  To obtain the systematics due to the 
imperfect background shape description, we vary the shape of the 
function describing the background: the parameters of the ARGUS 
function are varied within their errors; we also fix the shape to the 
one obtained from the generic MC simulation, or use a square root 
function instead of the ARGUS function.  The largest difference in the 
results is treated as the systematic uncertainty in the background 
shape description.  A possible contribution from inclusive $B\to 
  D_{s1}(2536)^+ X$ and $q\bar q\to D_{s1}(2536)^+ X$ production is 
checked by examining the $D_{s1}(2536)^+$ mass distribution in a 
$B\bar B$ MC simulation from which the signal has been removed, and in 
data in the $\Delta E-M_{\rm bc}$ sideband; no signal is found in both 
cases, so an upper limit on the yield is used as the systematic 
uncertainty due to the peaking background.  Another systematic 
uncertainty is from the errors in the fractions $R^\prime$ and 
$R^{\prime\prime}$ of the $D^{*0}$ and $D_{s1}(2536)^+$ subdecay 
modes, respectively.  These ratios are varied in the fit within their 
errors, and the difference in the fit results are assigned as a 
systematic uncertainty.  All the individual systematic errors, shown 
in Table~\ref{systematics}, are summed in quadrature.  Since the total 
systematic uncertainty is nearly symmetric, the maximum of the 
positive and negative errors is taken as the final systematic 
uncertainty.  The errors in the results are dominated by statistical 
uncertainties.
\begin{table}[htb]
  \begin{center}
	\caption{Relative systematic errors for the branching fractions
    	     and the ratio of the branching fractions, ${\cal B}(D_{s1}\to 
	     D^{*0}K^+)/{\cal B}(D_{s1}\to D^{*+}K^0)$, in \%.}
	\begin{tabular}{lcccc}
    \hline
    Source			   & ${\cal B}(D_{s1}\bar D^0)$
								   & ${\cal B}(D_{s1}D^-)$
										       & ${\cal B}(D_{s1}D^{*-})$
											  & $R^{\prime\prime}$ \\ 
    \hline
    Tracking            & $\pm9$    & $\pm9$    & $\pm9$   & $\pm3$ \\
    Particle ID         & $\pm3$    & $\pm3$    & $\pm3$   & $\pm1$ \\
    $\gamma/\pi^0$      & $\pm3$    & $\pm3$    & $\pm3$   & $\pm3$ \\
    $K_S$               & $\pm1$    & $\pm1$    & $\pm1$   & $\pm1$ \\
    MC resolution       & $\pm3$    & $\pm3$    & $\pm3$   & $\pm6$ \\
    BG shape            & $\pm4$    & $\pm4$    & $\pm4$   & $\pm3$ \\
    Peaking background  &$^{+0}_{-3}$ & $^{+0}_{-3}$ & $^{+0}_{-3}$ & $\pm3$ \\
    ${\cal B}(D^{(*)})$ & $\pm7$    & $\pm5$    & $\pm7$   & $\pm2$ \\
    $R^\prime$          & $\pm1$    & $\pm1$    & $\pm1$   & $\pm1$ \\
    $R^{\prime\prime}$  & $\pm1$    & $\pm1$    & $\pm1$   & $-$ \\
    $N(B\bar B)$        & $\pm1.5$  & $\pm1.5$  & $\pm1.5$ & $-$ \\
    \hline
    Total               & $\pm14$   & $\pm13$   & $\pm14$  & $\pm9$ \\
    \hline
  \end{tabular}
\label{systematics}
\end{center}
\end{table}

In summary, we report a measurement of the branching fractions for
the decays $B\to D_{s1}(2536)^+\bar D^{(*)}$, where $\bar D^{(*)}$ is
$\bar D^0$, $D^-$ or $D^{*-}$.  From a simultaneous fit to all $B$ and
$D_{s1}(2536)^+$ decay channels we measure ${\cal B}(B^+\to
D_{s1}(2536)^+\bar D^0)\times{\cal
  B}(D_{s1}(2536)^+\to(D^{*0}K^++D^{*+}K^0))
=(3.97\pm0.85\pm0.56)\times10^{-4}$, ${\cal B}(B^0\to D_{s1}(2536)^+
D^-)\times{\cal B}(D_{s1}(2536)^+\to(D^{*0}K^++D^{*+}K^0))
=(2.75\pm0.62\pm0.36)\times10^{-4}$ and ${\cal B}(B^0\to
D_{s1}(2536)^+D^{*-})\times{\cal
  B}(D_{s1}(2536)^+\to(D^{*0}K^++D^{*+}K^0))
=(5.01\pm1.21\pm0.70)\times10^{-4}$.  The ratio ${\cal B}(D_{s1}\to
D^{*0}K^+)/{\cal B}(D_{s1}\to D^{*+}K^0)$ is measured to be
$0.88\pm0.24\pm0.08$.  The first error is statistical and the second
one is systematic.  The obtained results are consistent within
errors with the previous measurements~\cite{ddk_babar}.

Using the latest measurements of the $B\to D^{(*)}D_{s(J)}^{(*)}$
branching fractions~\cite{pdg10} we calculate the ratios discussed
in~\cite{datta}:
\begin{eqnarray}
R_{D0} =\frac{{\cal B}(B\to DD_{s0}^*(2317))}
			 {{\cal B}(B\to DD_s)}=0.10\pm0.03, \nonumber \\
R_{D^*0}=\frac{{\cal B}(B\to D^*D_{s0}^*(2317))}
			 {{\cal B}(B\to D^*D_s)}=0.15\pm0.06, \nonumber \\
R_{D1} =\frac{{\cal B}(B\to DD_{s1}(2460))}
			 {{\cal B}(B\to DD_s^*)}=0.44\pm0.11, \nonumber \\
R_{D^*1}=\frac{{\cal B}(B\to D^*D_{s1}(2460))}
			 {{\cal B}(B\to D^*D_s^*)}=0.58\pm0.12. \nonumber
\end{eqnarray}
In addition, the same ratios are calculated for 
$B\to D^{(*)}D_{s1}(2536)^+$ decays using combined 
BaBar~\cite{ddk_babar} and current results:
\begin{eqnarray}
R_{D1^\prime} =\frac{{\cal B}(B\to DD_{s1}(2536))}
				   {{\cal B}(B\to DD_s^*)}=0.049\pm0.010,\nonumber\\
R_{D^*1^\prime}=\frac{{\cal B}(B\to D^*D_{s1}(2536))}
				   {{\cal B}(B\to D^*D_s^*)}=0.044\pm0.010.\nonumber 
\end{eqnarray}
In these calculations it is assumed that the decay modes 
$D_{s0}^*(2317)^+\to D_s^+\pi^0$ and 
$D_{s1}(2536)^+\to (D^{*0}K^++D^{*+}K^0)$ are dominant.

According to~\cite{orsay,datta}, within the factorization model and in 
the heavy quark limit, these ratios should be of order unity for the 
$D_{s0}^*(2317)$ and $D_{s1}(2460)$, whereas for the $D_{s1}(2536)$ 
they can be very small.  From the above ratios we can conclude that 
while the decay pattern of the $D_{s1}(2536)$ follows the 
expectations, the new $D_{sJ}$ states are either not canonical 
$c\bar s$ mesons, or this approach does not work for these particles.

We are grateful to A.~Datta for useful discussions. 
We thank the KEKB group for the excellent operation of the
accelerator, the KEK cryogenics group for the efficient
operation of the solenoid, and the KEK computer group and
the National Institute of Informatics for valuable computing
and SINET3 network support.  We acknowledge support from
the Ministry of Education, Culture, Sports, Science, and
Technology (MEXT) of Japan, the Japan Society for the 
Promotion of Science (JSPS), and the Tau-Lepton Physics 
Research Center of Nagoya University; 
the Australian Research Council and the Australian 
Department of Industry, Innovation, Science and Research;
the National Natural Science Foundation of China under
contract No.~10575109, 10775142, 10875115 and 10825524; 
the Ministry of Education, Youth and Sports of the Czech 
Republic under contract No.~LA10033 and MSM0021620859;
the Department of Science and Technology of India; 
the BK21 and WCU program of the Ministry Education Science and
Technology, National Research Foundation of Korea,
and NSDC of the Korea Institute of Science and Technology Information;
the Polish Ministry of Science and Higher Education;
the Ministry of Education and Science of the Russian
Federation and the Russian Federal Agency for Atomic Energy;
the Slovenian Research Agency;  the Swiss
National Science Foundation; the National Science Council
and the Ministry of Education of Taiwan; and the U.S.\
Department of Energy.
This work is supported by a Grant-in-Aid from MEXT for 
Science Research in a Priority Area (``New Development of 
Flavor Physics''), and from JSPS for Creative Scientific 
Research (``Evolution of Tau-lepton Physics'').


\begin{thebibliography}{99}
\bibitem{cc} The inclusion of charge-conjugate modes is implied
  throughout this Letter.
\end{thebibliography}
\end{document}